# A cross-correlation technique in wavelet domain for detection of stochastic gravitational waves

S.Klimenko, G.Mitselmakher, A.Sazonov
University of Florida

## Abstract

Stochastic gravitational waves (SGW) can be detected by measuring a cross-correlation of two or more gravitational wave (GW) detectors. In this paper we describe an optimal SGW search technique in the wavelet domain. It uses a sign correlation test, which allows calculation of the cross-correlation significance for *non-Gaussian data*. We also address the problem of *correlated noise* for the GW detectors. A method that allows calculation of the *cross-correlation variance*, when data is affected by correlated noise, is developed. As a part of the optimal search technique a robust estimator for detector noise spectral amplitude is introduced. It is not sensitive to outliers and allows application of the search technique to *non-stationary data*.

## 1 Introduction

Recently there is impressive progress in the development of gravitational wave (GW) interferometers [1-5]. One interferometer (TAMA, Japan) is collecting data and several more (LIGO, VIRGO and GEO) are about to start data taking. In the frame of an extensive scientific program, these interferometers will be used to search for stochastic gravitational waves (SGW). The SGW might be produced by processes in the early universe and by a large number of independent and unresolved GW sources [6-9]. It is exceptionally weak and a single detector can not distinguish the SGW from instrumental noise. However, if the SGW is correlated between several detectors, it can be detected using their cross-correlation. By integrating the cross-product of the detector output signals over a long period of time, one can expect to enhance the signal to noise ratio (SNR) if the instrumental noise is not correlated.

This technique for detecting of the SGW using two or more gravitational wave detectors was described in [10]. It uses a linear correlation test [11], which allows estimation of the significance of observed correlation if the instrumental noise is stationary and has a Gaussian distribution. A robust version of this technique, which effectively truncates detector sample values falling outside the central Gaussian-like part of the sample probability distribution, is described elsewhere [12,13]. In this paper (Section 2), a different cross-correlation technique is considered. It is based on the robust correlation test, which is much less compromised by the noise non-Gaussianity and non-stationarity. In Section 4 it is illustrated how this technique works with two GW detectors.

A correlated instrumental noise arising from the environment can be a serious problem for the cross-correlation techniques described above. A weak background from seismic events, power supplies and other environmental sources may result in statistically significant correlation between the GW detectors, which may be misinterpreted as the SGW or affect the SGW upper limit. The SGW's contribution to the cross-correlation depends on the relative orientation of the detectors. As suggested in [14], the contributions of the signal and the correlated noise can be estimated separately by changing the orientation of one of the detectors, which is, in this case, the resonant bar detector ALLEGRO [15] located in Louisiana State University (used in pair with the LIGO Livingston interferometer). Unfortunately, this method is not the ultimate solution of the problem. First, it is not applicable to the GW interferometers, which are permanently located. Second, the variance of the cross-correlation distribution can be very dependent on correlated noise. In Section 3 we address the problem of correlated noise and give a method, how to calculate the variance of the cross-correlation, if data is affected by correlated noise.

# 2 Robust correlation test

## 2.1 *Statement of the problem*

To characterize a correlation between two random variables *x* and *y*, which yield the data sets $\boldsymbol{x} = \{x_1, ..., x_n\}$ and $\boldsymbol{y} = \{y_1, ..., y_n\}$, usually the linear correlation coefficient *r* (or *Pearson's r*) is used [11]

$$r = \frac{\sum_i (x_i - \bar{x})(y_i - \bar{y})}{\sqrt{\sum_i (x_i - \bar{x})^2} \sqrt{\sum_i (y_i - \bar{y})^2}}, \quad (2.1)$$

where $\bar{x}$ and $\bar{y}$ are the sample mean values of the ***x*** and ***y*** respectively. When the correlation is known to be significant, the coefficient *r* is one conventional way to summarize its strength. However there is no universal way to compute the *r* distribution in the case of the null hypothesis, which is: *the variables x and y are not correlated*. In other words *r* is a poor statistic to decide whether a correlation is statistically significant or whether one observed correlation is significantly stronger then another if the data is not Gaussian.

To solve this problem, often a rank statistic [11] is used (non-parametric *Spearman's* test). It has a precisely known probability distribution function, which allows calculation of the significance of observed correlation. The rank correlation test (RCT) is almost as efficient[1] as the linear correlation test (LCT) and potentially it is a good choice to be used for the SGW search. However, in this paper the rank correlation is not discussed in details. The rank test is based on sorting algorithms, which are computationally intensive for large data sets. Instead, a robust correlation test (*sign test*) is considered, which is much simpler to use and easier to implement. Both correlation tests, rank and sign, can be used for the SGW search.

## *2.2 Description of the sign test*

Let us assume a data set ***x*** to be produced by a random process with zero median. In cases when the median is not zero, a new data set $\boldsymbol{x} - \hat{x}$ can be calculated, where $\hat{x}$ is the sample median of the ***x***. If the value of each $x_i$ is replaced with its sign $u_j$, the resulting first order statistics would be drawn from a known distribution function. It has zero mean and the $u_i$ will take on values +1 and –1. Given a second data set ***y***, we repeat the procedure, replacing each value $y_i$ by its sign $u_i$. We can introduce a new variable $s = uu$. Then the *sign correlation coefficient* is simply a sample mean of the data set $\boldsymbol{s} = \{u_1 u_1, ..., u_n u_n\}$

$$r = \bar{s} = \overline{uu}. \quad (2.2)$$

A value of $r$ near zero indicates that the variables $u$ and $u$ are uncorrelated. Assuming that the samples of ***s*** are statistically independent[2], the coefficient $r$ has a binomial distribution

$$P(n,r) = \frac{n}{2} \cdot \frac{2^{-n} n!}{m!k!}, \quad m = \frac{n}{2}(1-r), \quad k = \frac{n}{2}(1+r), \quad (2.3)$$

where $n$ is the number of data samples. This equation can be re-written using *Stirling's* approximation for factorials

$$P(n,r) \approx \sqrt{\frac{n}{2p(1-r^2)}} \cdot (1-r)^{-m}(1+r)^{-k}. \quad (2.4)$$

For large *n*, the $P(n,r)$ can be also approximated by a Gaussian distribution with variance *1/n*

$$P(n,r) \approx \sqrt{\frac{n}{2p}} \cdot \exp\left(-\frac{nr^2}{2}\right), \quad (2.5)$$

and the confidence level is given by the complimentary error function. Since, the number of samples n is typically large, the Gaussian approximation is used below in the text.

If the medians of the *x* and *y* variables could be know *a priori*, this test would be a *non-parametric* test. Since the medians are estimated from the data, the test depends on the errors of $\hat{x}$ and $\hat{y}$, and $r$ may have a systematic shift from its true mean. It can be shown that this dependence is very weak and therefore the test is robust. Indeed, the mean of $u$ (and $u$) is distributed with variance *1/n*. Respectively, its contribution to the mean of the $s$ distribution fluctuates with variance $2/n^2$, that usually is much less then the intrinsic variance of $r$ given by the test (*1/n*). We could say that for large *n* the *sign correlation test (SCT)* is *quasi non-parametric*.

## 2.3 Comparison of correlation tests

The SCT has been studied using Gaussian signals and noise with various distributions. The goal of the study was to estimate the SCT efficiency in comparison to the linear correlation test (LCT), which is one of the most efficient correlation tests. For comparison we also estimated the efficiency of the rank correlation test (RCT).

For the data sets, first, two time series $n_x$ and $n_y$ consisting on uncorrelated white noise were generated. We tried various noise probability distribution functions: Gaussian, Gaussian with tails, asymmetric Gaussian and uniform. Then a Gaussian signal *g* was added to both time series, so the data (***x*** and ***y***) is a sum of uncorrelated noise and correlated signal

$$x = n_x + g, \quad y = n_y + g. \quad (2.6)$$

The amplitude of the signal was relatively small. The signal to noise ratio (SNR) was less then 0.5, where the SNR was calculated as a ratio of standard deviations of the signal and the noise distributions.

To compare the tests, the correlation coefficients were calculated. Figure 1 shows the correlation coefficients as a function of SNR for different tests (LCT, RCT, SCT). The SCT efficiency was estimated by calculating the ratio $r/r$ for different types of noise. This ratio (*the sign correlation efficiency*) doesn't depend much on the signal to noise ratio and for Gaussian noise it is around 64%. In comparison, the corresponding ratio for the RCT is 95%.

---

[1] See discussion of the correlation test efficiency in section 2.3

[2] The case, when samples of ***s*** are correlated is discussed in Section 3.

Figure 2 shows the sign correlation efficiency for different types of noise.

Typically, if applied to the same data, the SCT will detect correlation with less significance then the LCT. When the detector noise is Gaussian, approximately 2.4 times more data for the SCT (1.1 for the RCT) needs to be collected in order to achieve the same confidence level as the LCT. However, for Gaussian noise with tails, which is a typical type of the detector noise, the SCT can have comparable or better efficiency then the LCT.

# 3 Correlated noise

## 3.1 Statement of the problem

In case of two GW detectors, let say *L* and *H*, the output of each detector is a time series, which is a mixture of the SGW signal (*h*) and noise (*n*)

$$x_{L,H}(t) = h_{L,H}(t) + n_{L,H}(t) \qquad (3.1)$$

Assuming no correlation between signal and noise, the cross-correlation *S* has a mean value

$$\langle S \rangle = \langle h_L h_H \rangle + \langle n_L n_H \rangle \quad (3.2)$$

and variance

$$V = \langle S^2 \rangle - \langle S \rangle^2 . \quad (3.3)$$

One can see, the noise term $\langle n_L n_H \rangle$ may bias the mean value of the cross-correlation. This is one problem, which is beyond the scope of this paper. We consider a different problem. Our goal is to calculate the variance of *S*, which could be affected by correlated noise as well as the mean value.

The idea of introducing the sign correlation test was to make the calculation of *V* independent on the detector noise model. In Section 2, *V* was calculated for any distribution functions of the data $x_L$ and $x_H$, assuming that samples of the sign correlation data set ***s*** are not correlated. It may not be true in the case of correlated noise. Below we introduce the noise *autocorrelation function* and modify the sign correlation test accordingly, in order to calculate the variance of *S* when data is affected by correlated noise.

## 3.2 Autocorrelation function

The sign correlation data set ***s*** can be considered as a time series generated by some random process $s(t)$. In general, there could be a correlation between samples of ***s***. The correlation is described by the autocorrelation function $a(t)$ of the process $s(t)$, which characterizes the correlation between samples of ***s*** separated by time $\tau$[3].

The function $a(t)$ depends on the correlation processes between the data sets. For a specific correlation process we will assume that $a(t \geq T_s) = 0$[4], where $T_s$ is the *correlation time scale*. Namely, any samples of ***s***, separated by time $t$ greater then $T_s$, are not correlated.

If the process $s(t)$ is a white noise, the samples of ***s*** are not correlated at any $t$. Then $T_s = \Delta t$, where $\Delta t$ is the sampling time interval and the function $a(t) = 0$, if $t > 0$. We define this situation as the *uncorrelated noise* case.

Different correlation processes due to instrumental noise may have different correlation time scales. Figure 3 shows the autocorrelation function for the cross-correlation of the LIGO Livingston (LLO) and Hanford (LHO) interferometer channels. One can see that the autocorrelation function is not zero when $t < 30\sec$ (due to the power correlated noise) and it is closer to zero for larger time scale. If the power lines are removed from the interferometer data, thus eliminating this source of correlated noise, the autocorrelation function is much closer to the one expected for uncorrelated data.

[3] We use the normalization *a(0)*=1.
[4] We assume no noise processes exist with an infinite correlation time scale.

One can see, that the conventional sign correlation test described above, in fact, uses the uncorrelated noise model. Applying the test, we assume that the autocorrelation function satisfies

$$a(0) = 1, \quad a(t > 0) = 0, \quad (3.4)$$

neglecting the second and higher order statistics of the process $s(t)$. In the scope of this model, the *probability distribution function* of the correlation coefficient (Equation 2.3) is calculated and then we check, if the observed correlation is consistent with this distribution. The conventional correlation test tells us if the data sets are correlated at time scale $T_s \geq \Delta t$ and the null hypothesis reads accordingly: *data sets are not correlated*.

When some correlated process is present in the data, a different null hypothesis should be used: *data sets are not correlated at time scale greater then some scale* $T_s$. Here we assume, that the data set ***s*** is generated by some correlated noise process $s_c(t)$ with the time scale $T_s$ and we know its autocorrelation function $a_c(t)$:

$$a_c(t < T_s) \neq 0, \quad a_c(t \geq T_s) = 0. \quad (3.5)$$

In fact, we introduce a model for correlated noise, which takes into account the second order statistics of the process $s_c(t)$ and the correlation test should tell us if the null hypothesis is consistent with this model.

How do we know the noise autocorrelation function? One practical way is to estimate it from the data. Given a cross-correlation data set ***s***, we can measure the autocorrelation function $a(t)$ quite well if $t < T_s << T$, where $T$ is the observation time. If the data is dominated by correlated noise with the time scale greater then the observation time, *a priori* knowledge of $a(t)$ should be used.

Below we describe how to calculate the probability distribution function of the correlation coefficient, if the data is contaminated with correlated noise. It is shown that the correlation coefficient has Gaussian distribution and we calculate its variance, if the assumption (3.5) about the correlated noise is used.

## *3.3 Variance of the correlation coefficient*

To calculate the variance of the correlation coefficient the following procedure is used. First, we divide the correlation data set ***s*** on intervals $T_s$ long. Then we form a slice $s_j$ by taking $j^{th}$ sample from each interval. The total number of slices is $m = T_s / \Delta t$. For each slice we can calculate a correlation coefficient $r_j$. Note, in the scope of the noise model, since $a_c(t > T_s) = 0$, the data in each slice $s_j$ is supposed to be a white noise and the coefficients $r_j$ assumed to be normally distributed (Equation 2.5) with the variances *m/n*. The correlation coefficient $r$ is given by

$$r = \tfrac{1}{m} \sum r_j, \quad j = 1,..,m. \quad (3.6)$$

In the case of correlated noise the correlation coefficients $r_j$ may not be statistically independent and the variance of $r$ is no longer *1/n*. Of course, we could select one of the coefficients $r_j$ to perform the correlation test, however it would not be optimal. The correlation between coefficients $r_j$ is described with the covariance matrix ***M(***$m \times m$)

$$M_{ij} = con(r_i, r_j) = \langle (r_i - r), (r_j - r) \rangle . \quad (3.7)$$

The elements of ***M*** can be described with the noise autocorrelation function $a_c(t)$

$$M \approx \frac{m}{n}\begin{bmatrix} 1 & b_1 & b_2 & ... \\ b_1 & 1 & b_1 & ... \\ b_2 & b_1 & 1 & ... \\ ... & ... & ... & ... \end{bmatrix}, \quad b_i = a_c(i\Delta t) + \frac{n-m}{n} \cdot a_c(T_s - i\Delta t),\ i = 1,...,m-1. \quad (3.8)$$

We can find a de-correlating transform ***U*** [16]

$$\tilde{r}_j = U_{ij} \cdot r_i , \qquad (3.9)$$

where ***U*** is a matrix of eigenvectors of the covariance matrix ***M.*** Since $r_i$ are normally distributed, the coefficients $\tilde{r}_i$ are statistically independent and normally distributed with the variances $L_{ii}$, where $L$ is the diagonal matrix of eigenvalues of ***M.*** Then the correlation coefficient and its variance are

$$\tilde{r} = \tfrac{1}{m}\sum \tilde{r}_j , \quad \mathrm{var}(\tilde{r}) = \tfrac{1}{m}\tfrac{1}{m}\sum \Lambda_{jj} . \quad (3.10)$$

When the time scale $T_s$ is large ($T_s / \Delta t >> 1$), we have to deal with a huge matrix ***M***. Therefore it makes sense to calculate the variance of the correlation coefficient given by the equation 3.6. Since the individual coefficients $r_i$ are normally distributed, the correlation coefficient $r$ is also normally distributed with the variance

$$\mathrm{var}(r) = \tfrac{1}{m}\tfrac{1}{m}\sum M_{ij} = \tfrac{1}{n} v . \quad (3.11)$$

Note, that ***v*** is the ratio of variances of $r$ calculated for two different noise models: correlated noise (3.5) and uncorrelated noise (3.4). When correlated noise is not present, ***v****=1*. Therefore the sum of the covariance matrix elements is an excellent measure of correlated noise. Measuring the autocorrelation function *a(t)* of the process $s(t)$ and comparing the sum

$$v(T_s) = 1 + \frac{2}{n}\sum_{i=1}^{T_s/\Delta t}(n-i)a(i\Delta t) . \quad (3.12)$$

with unity, the contribution from correlated noise at different time scales can be estimated. Figure 4 shows the variance ratio as a function of $T_s$ for the cross-correlation of the LHO (2k) and LLO (4k) interferometer output signals. One can see that the cross-correlation is dominated by the noise from power lines. First, the variance ratio rapidly increases and then saturates, indicating that there is a strong component of correlated noise with the time scale of the order of 50 seconds.

Defining the *correlation time* $\boldsymbol{T_c}$ as

$$T_c = \int_0^\infty a(t)dt , \quad (3.13)$$

if $T_s << T$, the cross-correlation variance satisfies

$$\mathrm{var}(\, r\,) = \frac{v}{n} \approx \frac{2T_c}{T}\,. \quad (3.14)$$

One can see that the variance shows the usual inverse dependence on the integration time. Also, if data is dominated by noise with the correlation time greater then the sampling interval, it will affect the variance and therefore the significance of the cross-correlation.

Since the above considerations hold for any correlation process $s(t)$, in order to calculate the cross-correlation variance, the same method can be applied to the correlation processes due to the SGW signal and the detector noise. As soon as the autocorrelation function of the combined process, which includes the uncorrelated noise, correlated noise and the SGW, is measured (or known *a priori*), the cross-correlation variance can be calculated.

# 4 Correlation in wavelet domain

## *4.1 Wavelet transforms*

The term wavelet is usually associated with a function $y \in L_2(R)$ [5] such that the translation and dyadic dilation of $y \in L_2(R)$ constitute an orthonormal basis of $L_2(R)$ [17,18]. For the discrete wavelet transform, the basis is

$$y_{mn}(t) = 2^{n/2} y\left(2^n t - m\right),\ n, m \in Z\,, \qquad (4.1)$$

where *the mother wavelet function* $y$ satisfies

$$\int y\,(t)dt = 0\,. \qquad (4.2)$$

The important property of wavelets is a time-frequency localization of their basis. It allows a time-frequency representation of data, similar to a windowed Fourier transform. The result of the wavelet transform of a time series *x(t)* is an array of wavelet coefficients $p_{mn}$, where *m* is the time index and *n* is the scale (or layer) index. Applied to wavelet data, the correlation can be estimated as a function of the layer index, which represents different frequency bands of the data. The wavelet coefficients in each layer can be considered as a time series with the sampling interval $\Delta t_n = 2^n \Delta t$.

Like for the original data *x(t)*, the probability distribution function of the wavelet coefficients *p(t)* is not actually known. However, considering the detail wavelet coefficients, their distribution usually has zero mean and is also symmetric, which makes it convenient to perform the sign transform in wavelet domain.

To apply the wavelet method to the SGW search, we modify the linear correlation test described in [9]. First (Section 4.2), we calculate the linear cross-correlation in the wavelet domain and then (Sections 4.3, 4.4) we apply the sign cross-correlation test.

## *4.2 Cross-correlation in wavelet domain*

The cross-correlation between two detectors is

$$S = \int_{-T/2}^{T/2} dt \int_{-T/2}^{T/2} dt' x_L(t') x_H(t) Q\left(t - t', \Omega_L, \Omega_H\right), \qquad (4.3)$$

where T is the integration time and $W_H$ ($W_L$) is the orientation of H (L) interferometer. The integration kernel $Q$ is selected to maximize the correlation due to the stochastic GW signal. Decomposing $x_L$ and $x_H$ in the wavelet domain with a discrete wavelet transform

$$x_L(t) = \sum_{k,l} p_{kl} y_{kl}(t) , \qquad x_H(t) = \sum_{n,m} q_{mn} y_{mn}(t) \qquad (4.4)$$

where $y_{ij}$ is the orthonormal basis of wavelet functions, we can rewrite 4.3

$$S = \sum_{nm} \sum_{k,l} p_{kl} q_{mn} I_{kl,mn} , \qquad (4.5)$$

$$I_{kl,mn} = \int_{-T/2}^{T/2} dt \int_{-T/2}^{T/2} dt' y_{kl}(t') y_{mn}(t) Q(t - t', \Omega_L, \Omega_H) . \qquad (4.6)$$

---

[5] Space of all square-integrable functions.

The double integral $l_{kl,mn}$ can be calculated in Fourier domain. Using the time localization of wavelet functions it follows that

$$I_{kl,mn} = \int_{-\infty}^{\infty} df \hat{y}_{kl}(f) \hat{y}^*_{mn}(f) Q(f, \Omega_L, \Omega_H) \qquad (4.7)$$

where $\hat{y}(f)$ and $Q(f, \Omega_L, \Omega_H)$ are the Fourier transforms of $y(t)$ and $Q(t, \Omega_L, \Omega_H)$ respectively. Note, if $Q(t - t') = d(t - t')$, due to the orthogonality of wavelet basis, it holds that $I_{kl,mn} = d_{km} d_{\ln}$, where $d_{km}$ ($d_{\ln}$) is Kronecker delta. For the optimal kernel, because of good localization of wavelet functions in time and frequency, usually most of the $I_{kl,mn}$ terms are equal to zero. Because of the frequency localization, we can omit terms with $l \neq n$ and rewrite the Equation 4.5 as follows

$$S \approx \sum_n S_n , \quad S_n = \sum_{k,m} p_{kn} q_{mn} I_{kn,mn} \qquad (4.8)$$

where $S_n$ is the cross-correlation for the $n^{th}$ layer. Since wavelet functions for each layer are produced by translation in time of the same function $y_n$, the Fourier integral for $I_{kn,mn}$ is

$$I_{kn,mn} = I_n(t) = \int_{-\infty}^{\infty} df \hat{y}_n(f) \hat{y}^*_n(f) Q(f, \Omega_L, \Omega_H) \exp(-j2pft) ., \qquad (4.9)$$

where $t$ is the time lag between coefficients $p$ and $q$ ($t = \Delta t_n (k - m)$). Assuming that the noise of each detector is much larger in magnitude then the SGW signal, the noise mean square for the $n^{th}$ layer is

$$s^2_{nL} = \frac{1}{N_n} \sum_{m=1}^{N_n} p^2_{mn} , \qquad s^2_{nH} = \frac{1}{N_n} \sum_{m=1}^{N_n} q^2_{mn} . \qquad (4.10)$$

where $N_n$ is the number of samples in the layer. Introducing the coefficient of linear correlation for layer $n$

$$r_n(t) = \frac{1}{N_n s_{nH} s_{nL}} \sum_{m=1}^{N_n} p_n(m\Delta t_n) q_n(m\Delta t_n - t) , \qquad (4.11)$$

the cross-correlation sum $S$ calculated in wavelet domain is

$$S = \sum_n N_n s_{nL} s_{nH} \sum_t I_n(t) \cdot r_n(t) = \sum_{n,t} N_n w_n(t) r_n(t) . \qquad (4.12)$$

So far the cross-correlation $S$ is equivalent to the result obtained in the Fourier domain [9]. Similarly we can use the optimal kernel to calculate the integrals $I_n(t)$ and hence $w_n(t)$

$$Q(f,\Omega_L,\Omega_H) = l\,\frac{|f|^{-3}\,\Omega_{GW}(f)\Gamma(f,\Omega_L,\Omega_H)}{P_L(f)P_H(f)}, \qquad (4.13)$$

where $\Omega_{GW}(f)/f$ is proportional to the GW differential energy density, $g$ is the detector overlap reduction function [8], $P_L$, $P_H$ are the spectral densities of the detector noise and $l$ is the normalization constant. Then the full expression for the $w_n(t)$ is

$$w_n(t) = \int_{-\infty}^{\infty} df |y_n(f)|^2 |f|^{-3} \Omega_{GW}(f) g(f,\Omega_L,\Omega_H) \frac{s_{nL}}{P_L(f)} \cdot \frac{s_{nH}}{P_H(f)} \exp(-j2pft). \; (4.14)$$

The Equation 4.12 shows that the cross-correlation S is a weighted sum of the linear correlation coefficients $r_n(t)$. For a pair of GW detectors, the optimal weight coefficients can be calculated for a selected SGW model ($\Omega_{GW}(f)$) using an appropriate noise spectral density estimator. However, one question remains: What is the probability distribution function of the correlation coefficients $r_n(t)$ if the detector noise is non-Gaussian and/or correlated?

To solve this problem, the sign correlation test is used. The problem of the correlated noise can be solved, by using the noise autocorrelation function approach described in Section 3.

## 4.3 *Sign correlation in wavelet domain*

To apply the sign correlation test, we calculate the cross-correlation $S_s$ similar to the one given by Equation 4.12

$$S_s = \sum_i N_i w_i r_i, \quad i = (n,t), \quad (4.15)$$

where $w_i$ are the optimal coefficients, which we need to find, and $r_i$ are the sign correlation coefficients. Here for simplicity, one index $i$ is used instead of two indexes $n$ and $t$. The variance of $S_s$ is

$$V_s = \sum_i N_i w_i^2 v_i \,. \quad (4.16)$$

To find the optimal coefficients, we should maximize the signal to noise ratio (SNR)

$$\frac{\partial(SNR^2)}{\partial w_i} = \frac{\partial(m^2/V_s)}{\partial w_i} = 0, \quad (4.17)$$

where $m$ is the mean value of $S_s$ for stochastic gravitational waves with strength $W$[6]. Since the expectation values of the sign correlation coefficients are proportional to $W$ ($\bar{r}_i = l_i\Omega$) and

$$m = \Omega \sum_i N_i w_i l_i, \quad (4.18)$$

we can rewrite the equation 4.17 as follows

$$l_i \sum_j N_j w_j^2 v_j = w_i v_i \sum_j N_j w_j l_j \,. \quad (4.19)$$

The solution of this equation is $w_i = l_i / v_i$ and the optimal cross-correlation is

$$S_s = \sum_i N_i \frac{l_i}{v_i} r_i \,. \quad (4.20)$$

To find the coefficients $l_i$, a detail simulation of the SGW signal can be used. It should take into account the detector response functions and their overlap reduction function.

---

[6] Without loss of generality we assume $W(f)$=const.

Adding the simulated signal with known strength $\Omega$ to the detector output signals, we can find $\bar{r}_i$ and therefore $l_i$.

Also the coefficients $l_i$ can be calculated by using the optimal coefficients obtained in the previous section. Since the expectation values of the linear correlation coefficients are $\bar{r}_i = w_i\Omega$, with $w_i$ given by Equation 4.14, the coefficients $l_i$ can be found as

$$l_i = w_i \frac{\bar{r}_i}{\bar{r}_i} = w_i e_i\,, \quad (4.21)$$

where $e_i$ is the efficiency of the sign correlation test (see Section 2.3). Then the optimal coefficients for the sign cross-correlation sum are

$$w_i = w_i \frac{e_i}{v_i}\,. \quad (4.22)$$

It is much easier to obtain $e_i$ from the simulation then $l_i$, because they do not depend on the SGW model, detector responses and the detector overlap reduction function, which are taken into account by the coefficients $w_i$. For example, by adding Gaussian noise into the detector output, the same for both detectors, the efficiency $e_i$ can be calculated as the ratio of the sign and linear correlation coefficients.

It can be seen that the signal to noise ratio

$$SNR_s = \Omega\sqrt{\sum_i N_i \frac{w_i^2 e_i^2}{v_i}}\,, \quad (4.23)$$

is proportional to $\Omega\sqrt{T}$, because all $N_i$ are proportional to the total number of samples and therefore to $T$. Note, if the detector noise is Gaussian (all $e_i$ =0.64) and uncorrelated (all $v_i$=1), for the same data, the SNR obtained with the sign correlation test is by 64% smaller then the SNR for the linear correlation test.

## 4.4 Robust spectral amplitude

We can include the sign correlation efficiency $e_n$, which should not depend on the lag time $t$, in the expression for $w_n(t)$ (Eq.4.14) and introduce the *robust noise spectral amplitude*

$$A_I(f) = \frac{P_I(f)}{s_{nI}\sqrt{e_n}}\,, \qquad I=L,H\,. \qquad (4.24)$$

Then the optimal weight coefficients $\tilde{w}_n(t)$ are

$$\tilde{w}_n(t) = \int_{-\infty}^{\infty} df \left|y_n(f)\right|^2 |f|^{-3} \frac{\Omega_{GW}(f)\cdot g(f,\Omega_L,\Omega_H)}{A_L(f)\cdot A_H(f)} \exp(-j2pft)\,. \quad (4.25)$$

One can see that only the amplitudes $A_L$ and $A_H$ should be estimated from the experimental data to calculate the optimal weight coefficients.

The robustness of amplitudes $A_I$ can be shown on the example of Gaussian noise with tails. For simplicity we used a white noise, then $P_I(f) = const$. The tail contribution can be characterized by the ratio of the noise variance $s_n^2$ and the variance of its Gaussian core $s_g^2$, which remained constant. The value of spectral density $P$ is proportional to the total noise variance $s_n^2$ and it increases, when the tail contribution increases (see Table 1).

| $s_n^2 / s_g^2$ | P, a.u. | A, a.u. |
|---|---|---|
| 1.0 | 0.45 | 0.0266 |
| 2.1 | 0.94 | 0.0274 |
| 5.4 | 2.40 | 0.0273 |

Table 1. Dependence of the noise spectral density *P* and robust noise amplitude *A* on tail contribution.

At the same time one can see that the amplitude *A* remains constant. It makes calculation of the optimal weight coefficients more robust if the noise tails are non-stationary.

# 5 Measurement of *W*

As it was shown in the previous sections, the sign cross-correlation is normally distributed with the variance $V_s$ (Equation 4.16). From here, given a measurement $S_s$, we can calculate its significance level

$$SL = \frac{1}{2} erf\left(\frac{S_s}{\sqrt{V_s}}\right).$$

If the measurement is not statistically significant, following the procedure described in [19], the cross-correlation upper limit $\tilde{S}_s$ can be calculated. The mean value of the cross-correlation for stochastic gravitational waves with strength $W$ is

$$m = \Omega \sum_i N_i w_i v_i = \Omega V_s,$$

where $w_i$ is given by Equation 4.22. Then, given either the measured value of the cross-correlation $S_s$ or its upper limit $\tilde{S}_s$, we can estimate Ω ($\Omega = S_s / V_s$) or the upper limit on Ω ($\tilde{\Omega} = \tilde{S}_s / V_s$).

# 6 Conclusion

The SGW signal should manifest itself in the cross-correlation of several detectors. In this paper we considered a) the problem of estimation of the cross-correlation significance if the detector noise is not Gaussian and b) the problem of correlated noise.

Different correlation techniques can be used to calculate the detector cross-correlation. We compared the linear correlation test with the sign and rank tests, which allow us to calculate the variance of the cross-correlation if the detector noise is non-Gaussian. The rank correlation is a non-parametric test and it is almost as efficient as the linear correlation test. It is based on sorting algorithms and could be quite time consuming for large data sets. The sign correlation test is robust and simple to use, however it may be less efficient in comparison to the linear and rank correlation tests.

We addressed the problem of correlated noise and described a method how to calculate the cross-correlation and its variance, if data is affected by correlated noise. We also suggested a method how to estimate the amount of correlated noise in the data.

Using the sign correlation test, we introduced an optimal cross-correlation technique in the wavelet domain. It allows calculation of the cross-correlation coefficients for different wavelet scales as functions of lag time, which is an important signature of the correlation process. The optimal cross-correlation is calculated as a weighted sum over the correlation coefficients, where the weight coefficients are calculated for a specific SGW model, taking into account the power spectral densities of the noise and the detector overlap reduction function. As a part of the method we introduced a spectral density estimator, which allows a robust estimation of the noise spectral density, if the detector noise has outliers.

## Acknowledgments

We are grateful to Bruce Allen for very useful suggestions, which considerably improved the paper, and Bernard Whiting for comments and suggestions regarding the text. We also thank our REU student Jason Castiglione, who helped with calculations involving the rank and sign correlation tests. This work was supported by NSF grant PHY-0070854.

# Figures

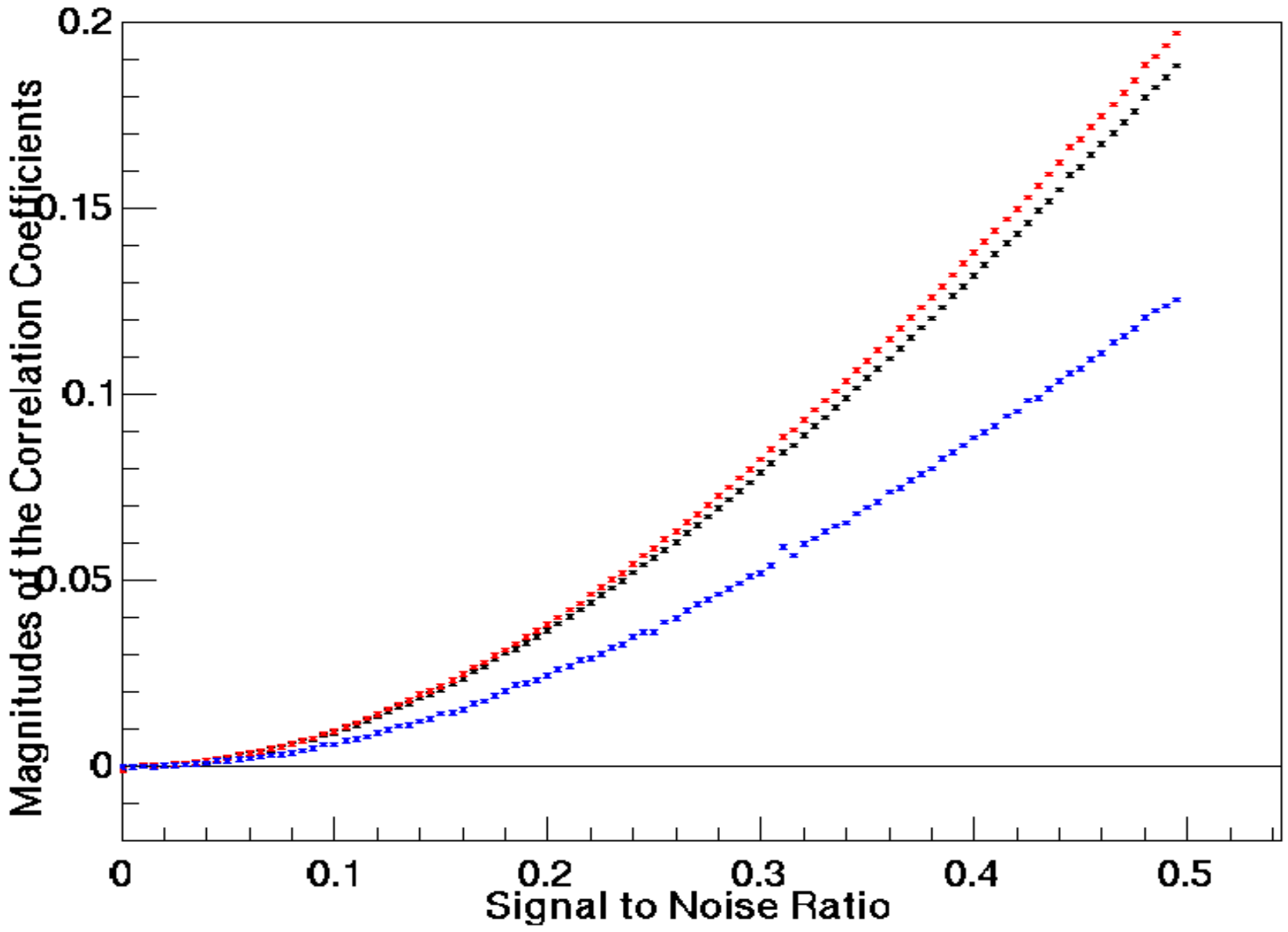

Figure 1. Correlation coefficients as a function of SNR for Gaussian noise: LCT(up), RCT (middle), SCT (down).

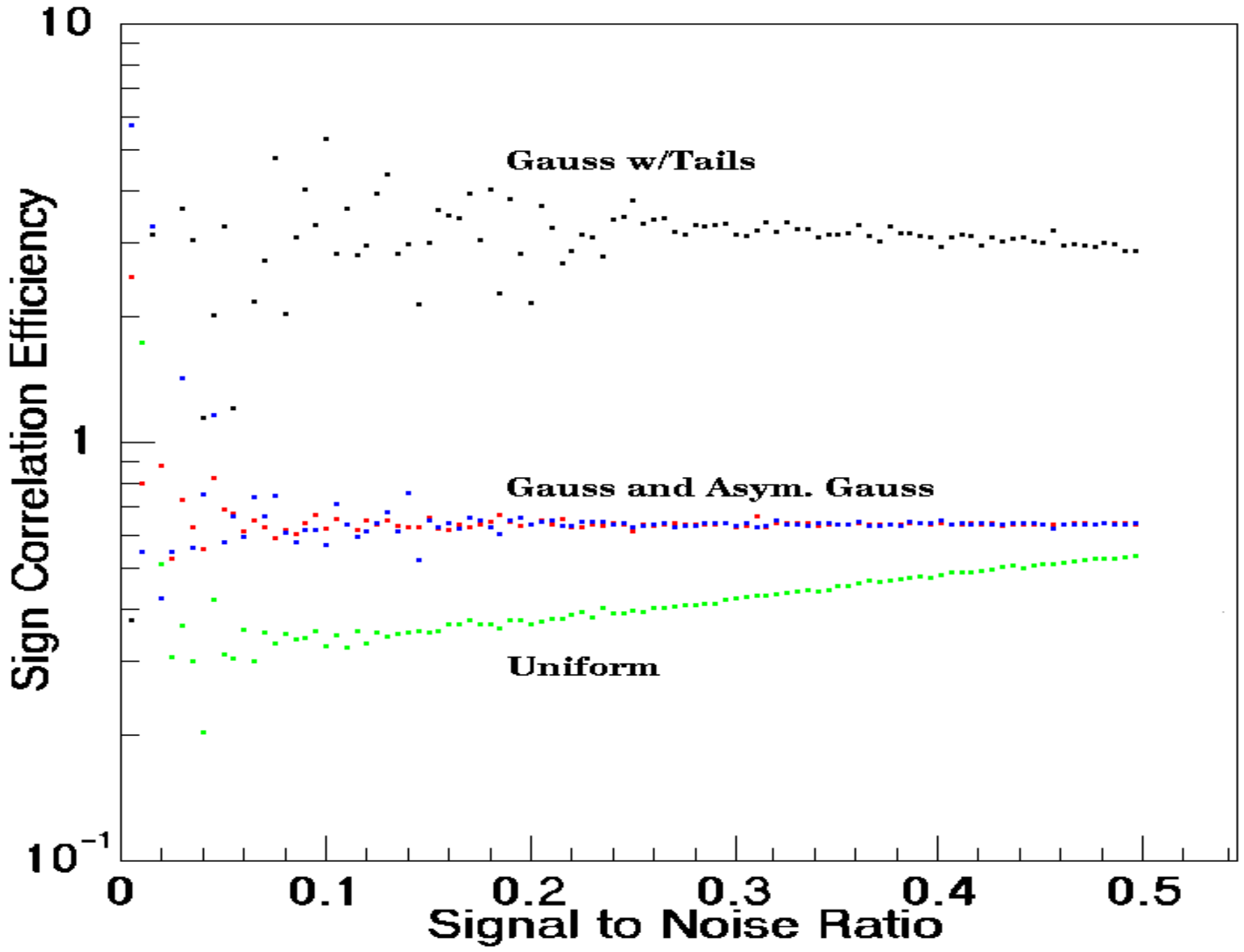

Figure 2. Sign correlation efficiency as a function of SNR for different noise distributions.

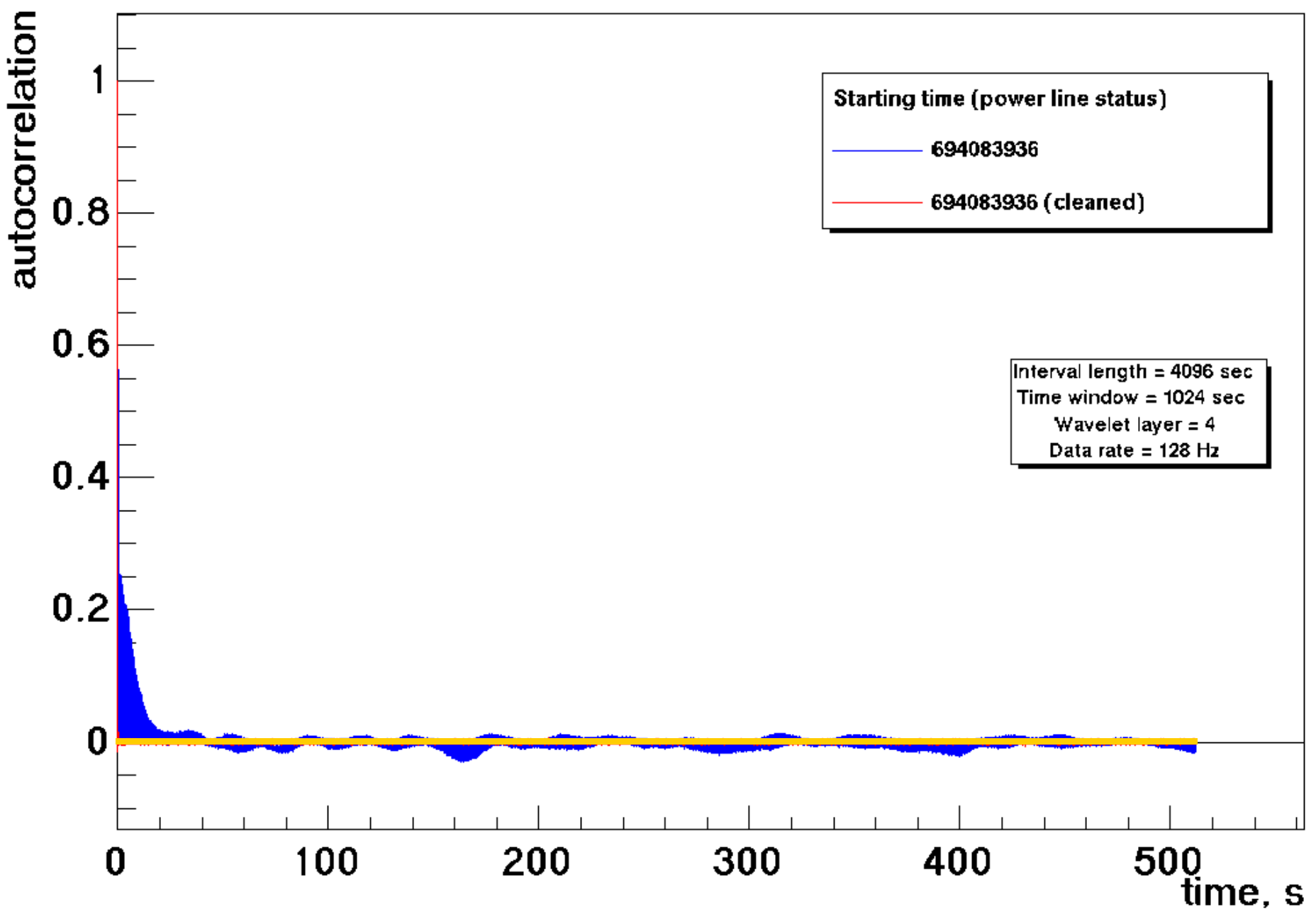

Figure 3. Autocorrelation function of LIGO data representing the frequency band of 32-64 Hz: blue (dark) – raw data, yellow (light) – power lines (with fundamental frequency 60 Hz) are removed.

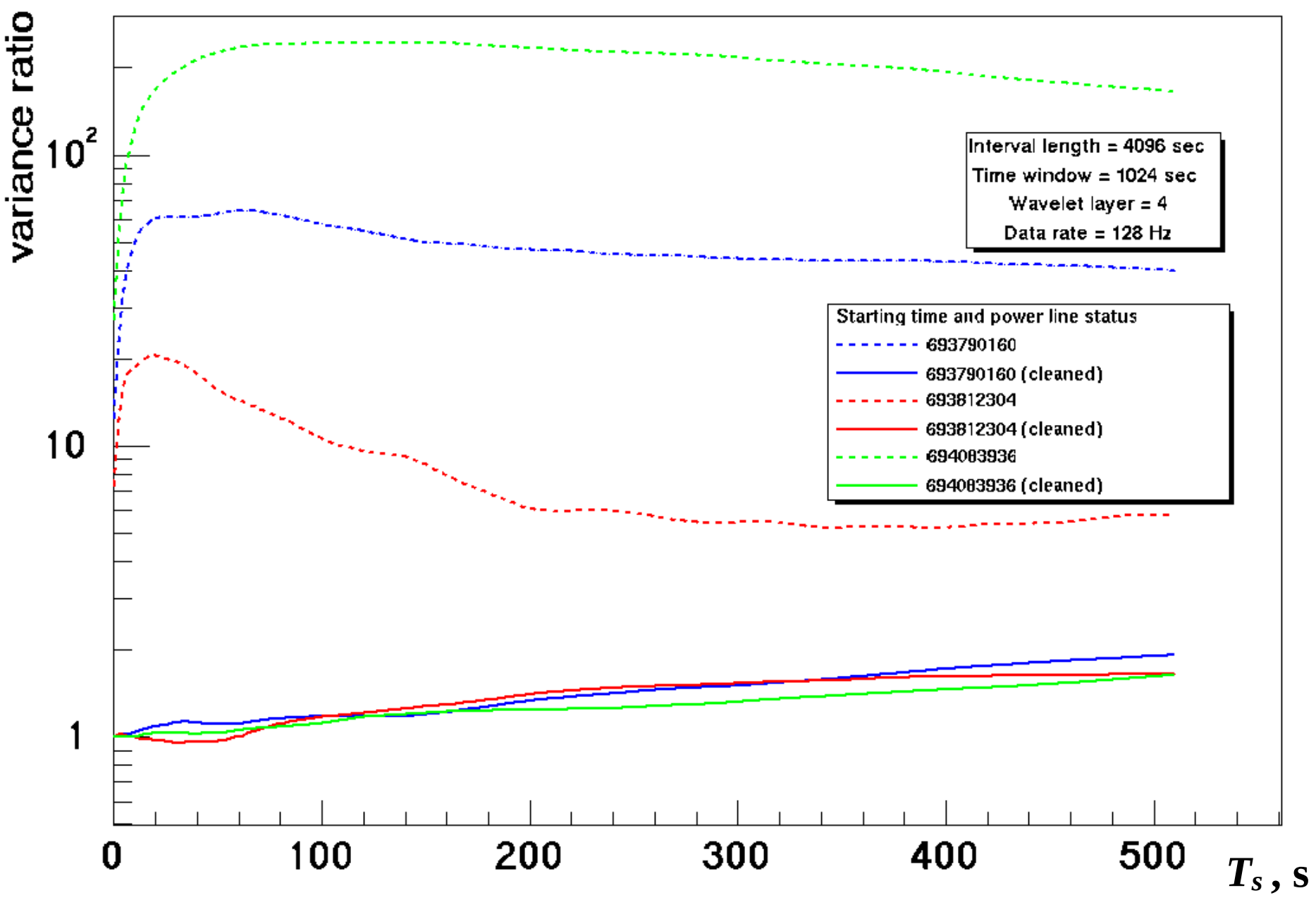

Figure 4. Variance ratio as a function of $T_s$ for 3 samples of LIGO data representing the frequency band of 32-64 Hz: dashed – raw data, solid - power lines (with fundamental frequency 60 Hz) are removed.